\documentclass[%
reprint,
superscriptaddress,
 amsmath,amssymb,
 aps,
]{revtex4-1}

\usepackage{graphicx}
\usepackage{dcolumn}
\usepackage{bm}

\usepackage{cmap} 
\usepackage[T2A]{fontenc} 
\usepackage[utf8]{inputenc} 
\usepackage[english]{babel} 
\usepackage{multirow}
\usepackage[warn]{mathtext}
\usepackage{amssymb}
\usepackage{ dsfont }
\usepackage{ textcomp }
\usepackage{ mathrsfs }

\renewcommand{\epsilon}{\ensuremath{\varepsilon}}
\renewcommand{\phi}{\ensuremath{\varphi}} 
\renewcommand{\kappa}{\ensuremath{\varkappa}}

\renewcommand{\geq}{\ensuremath{\geqslant}}

\usepackage{textcomp} 
\usepackage{indentfirst} 
\usepackage{amsmath} 
\usepackage{graphicx} 
\DeclareGraphicsExtensions{.pdf,.png,.jpg}
\usepackage{pgfplots}
\pgfplotsset{compat=1.13}

\usepackage{hyperref}
\hypersetup{
    colorlinks=true,
    linkcolor=blue,
    filecolor=black,      
    urlcolor=black,
    citecolor= violet
}

\renewcommand{\bibname}{References}

\usepackage[nottoc]{tocbibind}

\begin{document}

\makeatletter
\newcommand{\settitle}{\maketitle}
\makeatother

\title{Simulation of the five-qubit quantum error correction code on superconducting qubits}

\author{I. A. Simakov}
\affiliation{Moscow Institute of Physics and Technology, 141701 Dolgoprudny, Russia}
\affiliation{Russian Quantum Center, 143025 Skolkovo, Moscow, Russia}
\affiliation{National University of Science and Technology ``MISIS'', 119049 Moscow, Russia}
\author{I. S. Besedin}
\affiliation{Russian Quantum Center, 143025 Skolkovo, Moscow, Russia}
\affiliation{National University of Science and Technology ``MISIS'', 119049 Moscow, Russia}
\author{A. V. Ustinov}
\affiliation{Russian Quantum Center, 143025 Skolkovo, Moscow, Russia}
\affiliation{National University of Science and Technology ``MISIS'', 119049 Moscow, Russia}
\affiliation{Physikalisches Institut, Karlsruhe Institute of Technology, 76131 Karlsruhe, Germany}

\date{\today}

\begin{abstract}
Experimental realization of stabilizer-based quantum error correction (QEC) codes that would yield superior logical qubit performance is one of the formidable task for state-of-the-art quantum processors. A major obstacle towards realizing this goal is the large footprint of QEC codes, even those with a small distance. We propose a circuit based on the minimal distance-3 QEC code, which requires only 5 data qubits and 5 ancilla qubits, connected in a ring with iSWAP gates implemented between neighboring qubits. Using a density-matrix simulation, we show that, thanks to its smaller footprint, the proposed code has a lower logical error rate than Surface-17 for similar physical error rates. We also estimate the performance of a neural network-based error decoder, which can be trained to accommodate the error statistics of a specific quantum processor by training on experimental data.
\end{abstract}

\settitle

\section{Introduction}

The realization of quantum error correction codes is an essential step towards scalable quantum computation. Stabilizer codes \cite{ShorQEC, DiVincenzoShor,GottesmanPhd, SteaneQEC, SteaneQEC2} encode ``logical'' qubits in ``physical'' qubits. If the error rate of a physical qubit is less than a certain threshold level, then the composite logical qubit will have a lower error rate \cite{knill1996threshold}. Among the plethora of different QEC codes that have been proposed, topological QEC codes, such as surface codes \cite{BravyiKitaev, SurfaceCodes, XZZXSurfaceCode} and color codes \cite{Bombin2006, Bombin2007} particularly stand out. The topological property means that there exists an arrangement of the physical qubits on a finite-dimensional lattice such that only local operators need to be measured each error correction cycle. This property is important for practical realization on a number of physical platforms, such as superconducting qubits, where high-fidelity two-qubit gates are often only available between neighboring qubits.
As a consequence, many recent experiments and near-term proposals for stabilizer codes on superconducting qubits involve small-scale surface codes \cite{ibmart, DiCarlo2017, DiCarloLeakage2020, googleRepetitveCode, andersen2020repeated, DiCarlo2021logical}. 

Performance of error correction codes depends not only on the frequency of errors, but also on the type of errors. Simulation of the performance of different QEC codes relies on an accurate and physically motivated error model, which can be computationally expensive, but still viable for small-scale circuits, such as Surface-17 \cite{OBrien2017,LowDistanceSurfaceCodes}. For larger codes, approximate error models have been proposed \cite{Bravyi2018, SurfaceCodeDecoherence}.

Furthermore, the problem of an optimal quantum error decoder remains relevant \cite{Fowler2012, varsamopoulos2019decoding}. One of the most reliable conventional decoding algorithms is the minimum weight-perfect matching (MWPM) algorithm, which shows good performance for different codes and in particular for surface codes \cite{OBrien2017, LowDistanceSurfaceCodes, das2020scalable}. Accuracy of the MWPM decoder relies on an approximate error model, which may or may not be justified for real devices. The key advantage of MWPM decoders is speed, as the runtime of the decoder scales polynomially with the size of the circuit. This makes MWPM decoders good candidates for large-scale error correction, but questionable for early small-scale codes with qubits and gate fidelities barely over the error correction threshold.
Artificial neural network-based decoders are another promising type of decoders \cite{sheth2020neural, baireuther2019neural}. Their runtime and accuracy varies, and thanks to their architecture they can accommodate error statistics directly from measured data \cite{baireuther2018machine}, demonstrating better logical error rates compared to MWPM.

Even though surface codes and color codes have numerous important properties that make them good candidates for scalable universal quantum computation, they are by far not the most efficient codes in terms of code distance per physical qubit. Unfortunately, the stabilizers in these more efficient codes have a complex structure, and designing a circuit to measure them with high fidelity has proven to be a challenging task. The smallest possible distance-3 code that corrects all possible single-qubit errors is the 5-qubit ZXXZ code. Schuch and Siewert made a proposal of a quantum circuit that utilizes 9 qubits connected in a ring topology with iSWAP gates available on neighboring qubits \cite{schuch2003natural} implementing this code.

In this paper, we propose a quantum circuit for the 5-qubit distance-3 QEC code appropriate for experimental realization on a ring of 10 superconducting qubits. We estimate the upper bound performance of the code using a time-domain density-matrix simulation of the circuit under a realistic noise model and compare its performance to the surface code of the same distance. For forthcoming experiments, we present a neural network decoder based on long short-term memory architecture, to be trained on only experimentally available data.

\begin{figure*}[t]
    \center{\includegraphics[width=\linewidth]{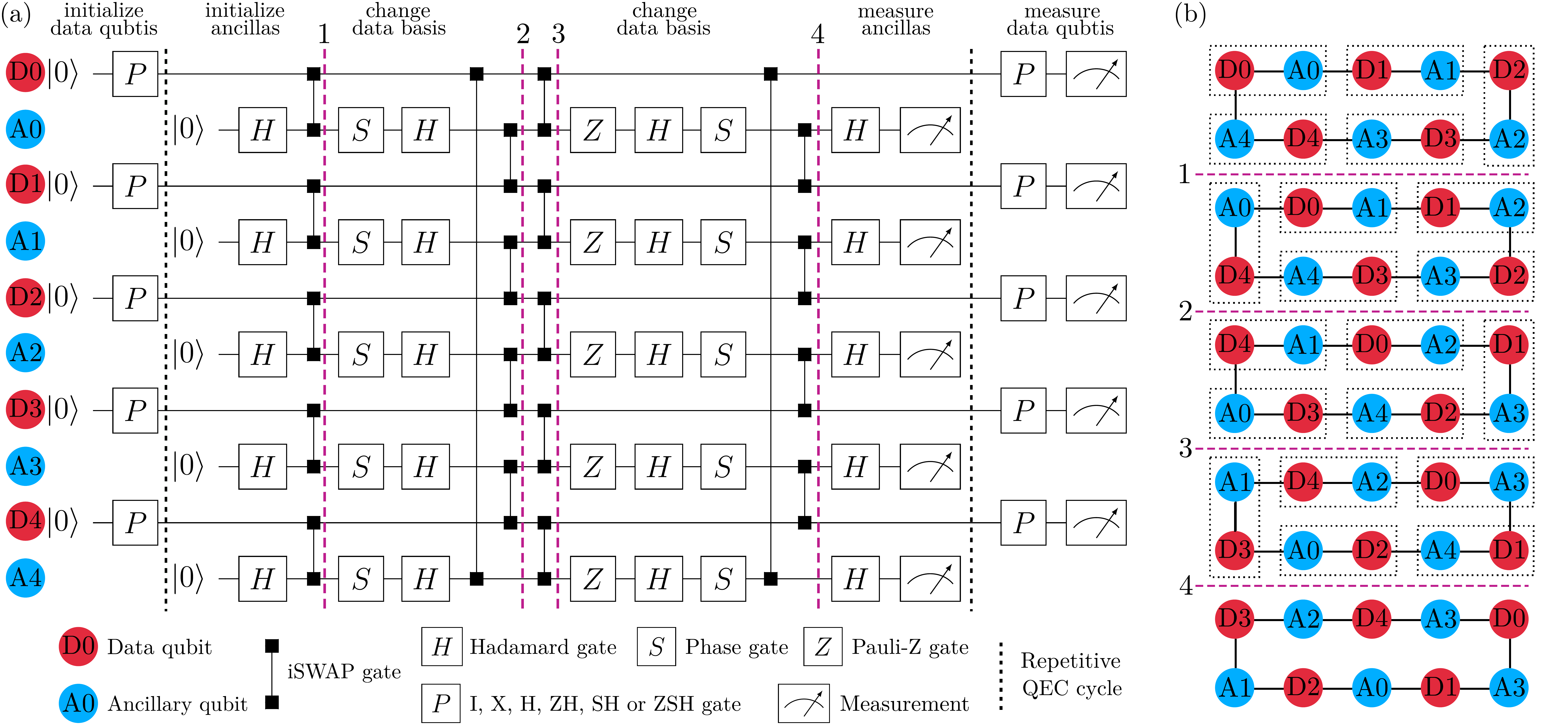}}
    \caption{a) The quantum circuit for the simulated state preservation experiment for the five-qubit code. The scheme is divided into three stages: initialization of the data qubits, repetitive error correction cycles, and final measurement of the data qubits. During the correction cycle there are four series of simultaneous iSWAP gates between the neighboring qubits, highlighted with the dashed purple lines. b) Schematic layout of computational qubit positions during the correction cycle. Qubits that are paired in the same iSWAP gate at each step are outlined with a dashed rectangle. The numbered dashed purple lines correspond to the ones in panel (a).}
    \label{circuit-10}
\end{figure*}

\section{Methods}

In this section we describe the quantum circuit of the five-qubit code and give the details of the simulated experiment.

\subsection{Quantum circuit}

The five-qubit QEC code encodes 1 logical qubit in 5 data qubits, protecting it against arbitrary single-qubit errors. Its four syndrome generators are
\begin{equation}
    \begin{aligned}
        g_0 &= Z_0 X_1 X_2 Z_3 I_4 \\
        g_1 &= I_0 Z_1 X_2 X_3 Z_4 \\
        g_2 &= Z_0 I_1 Z_2 X_3 X_4 \\
        g_3 &= X_0 Z_1 I_2 Z_3 X_4
    \end{aligned}
\end{equation}
and Pauli operators on a logical qubit are expressed by the following single-qubit operators:
\begin{equation}
\label{eq:logical_operators}
    \begin{aligned}
        X_L &= X_0 X_1 X_2 X_3 X_4 \\
        Y_L &= Y_0 Y_1 Y_2 Y_3 Y_4 \\
        Z_L &= Z_0 Z_1 Z_2 Z_3 Z_4,
    \end{aligned}
\end{equation}
where $X_i$, $Y_i$, $Z_i$, $I_i$ are Pauli operators applied to the $i$-th qubit. 


The quantum circuit for this experimental procedure is presented in Fig.~\ref{circuit-10}a. It consists of ten physical qubits, five of which are data qubits and the rest are ancillary qubits used to extract the error syndromes. The circuit operation can be divided into three stages: data qubit initialization, $k$ repetitions of the error correction cycle, and measurement of the data qubits. 

In the initialization stage, we assume that each qubit can be prepared in the $|0\rangle$ state. To initialize other states, a single-qubit gate denoted as $P=X, H, ZH, SH \text{ or } ZSH$ can transform the $|0\rangle$ state into  $|1\rangle, |+\rangle, |-\rangle, |i+\rangle, \text{ or } |i-\rangle$, respectively.

The syndrome measurement stage consists of $k$ cycles. Each cycle starts with the initialization of the ancillary qubits. The most important parts of the cycle are four series of iSWAP gates that entangle neighboring qubits. The action of the iSWAP gate on two qubits can be decomposed into a swap of their computational states, a CZ gate, and single-qubit $S$ gates on both qubits. Thereby, it appears that data qubits move clockwise in the ring and ancillas move counterclockwise, as it is illustrated in Fig~\ref{circuit-10}b. In this cartoon we schematically show the position of computational qubits before and after each series of iSWAP gates. Qubits that are paired in the same iSWAP gate at each step are outlined with a dashed rectangle. At the end of the correction cycle we measure all ancillas and reset them for the next error correction cycle.
It should be noted that one ancilla measurement is redundant. The extra syndrome $g_4 = X_0 X_1 Z_2 I_3 Z_4$ can be expressed as the product of the other syndromes.

Finally, after $k$ QEC cycles we act on the data qubits with the same gate $P$ as in the initializing procedure and projectively measure them in the $Z$ basis.

In theory the five-qubit code, as the majority of other QEC codes, protects the logical state against one arbitrary single-qubit error that happens to data qubit between the correction cycles. It is vulnerable for measurement errors, initialization errors, multi-qubit errors and imperfection of gates. But, fortunately, the impact of these unwanted errors can be significantly reduced by using a smart decoding algorithm that will have access to all measured data over several sequential correction cycles.

\subsection{Simulated experiment}

To quantify the performance of the five-qubit code we focus on simulation of the state preservation experiment. The experimental procedure is the following: we initialize data qubits in one of the logical states $|0\rangle, |1\rangle, |+\rangle, |-\rangle, |i+\rangle, |i-\rangle$, execute the syndrome measurement gates for $k$ cycles, and then measure the final state.

For the simulation we calculate time-domain density-matrix evolution and use operator-sum representation for gates and error channels. In our model we take into account amplitude and phase damping during qubit idling and gates, and consider the phenomenological depolarizing effect and stochastic errors of the rotation gate phase in order to resemble gates implemented in \cite{timeerrorcite, fluxnoisecite1, fluxnoisecite2}. Also, we take into account measurement errors that are similar to the ones observed in a real experiment \cite{measerrorcite}. The main time parameters used in the simulation are given in Table~\ref{time_params}. We calculate the gate fidelities with the formula \cite{nielsen2002simple}
\begin{equation}
    F_{\text{gate}} = \frac{\text{Tr}(R_{\text{ideal}}^\dagger R) + d}{d(d+1)},
    \label{gate_fid}
\end{equation}
where $d=2$ for single-qubit gates, $d=4$ for two-qubit gates, and $R_{\text{ideal}}$ and $R$ are Pauli Transfer matrices corresponding to the actions of the ideal and simulated gates. The resulting fidelities in our simulation are 0.9995 for single-qubit rotation gates and 0.998 for two-qubit iSWAP gates. More detailed information about the error models is given in the Appendix~\ref{appendix:errors}.

\begin{figure*}[t]
    \center{\includegraphics[width=\linewidth]{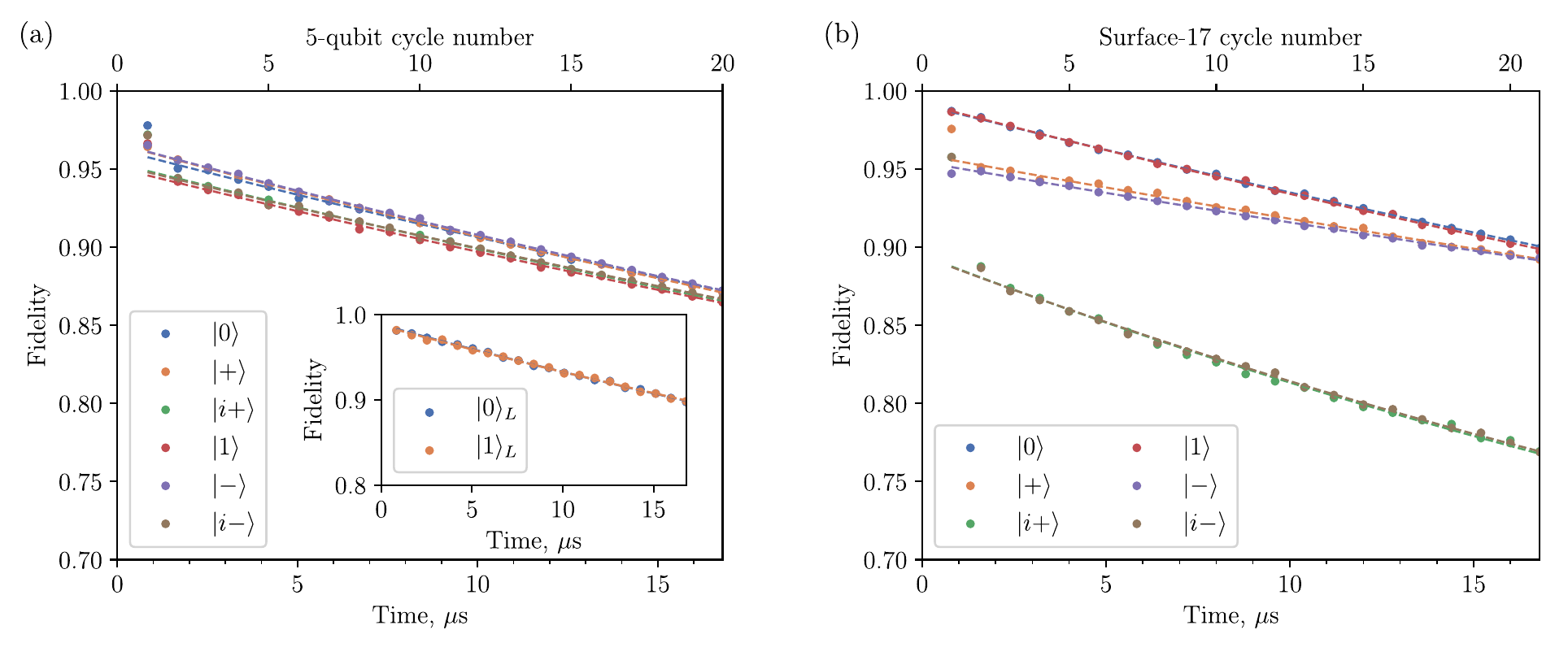}}
    \caption{Comparison of the ideal (upper bound) decoder performance of (a) the five-qubit code and (b) Surface-17 for different initial states of the data qubits. The dashed curve is the least-squares fit of the corrected state fidelity with Eq. ~\eqref{fid_approx}. The inset plot in figure (a) shows the decay of the exact logical states $|0\rangle$ and $|1\rangle$ of the five-qubit code.}
    \label{dif_init_states}
\end{figure*}

\section{Results}

In the current section we estimate the best possible performance of the five-qubit code, compare it to surface-17 code and present the logical error rate dependence on the qubit coherence. Further, we suggest a budding decoder based on an artificial neural network and test it on a problem of an arbitrary initial state correction.

\subsection{Upper bound performance}

The decoder is a function that gets the ancilla qubit measurement results as an input, and returns a unitary operator $U$ that should be applied to the final logical state to reconstruct the initial state as its output. The ideal decoder returns the unitary operator that yields the best possible fidelity of the final state with respect to the pure initial state. This fidelity cannot exceed the largest eigenvalue of the final density matrix, that gives a decoder-independent upper bound for the error correction code performance.

Using the density matrix of the final state of the 5 data qubits $\rho$ we construct the density matrix of the logical qubit $\rho_L$:
\begin{equation}
    \rho_L = \cfrac{I +
    \text{Tr} (X_L \rho) X +
    \text{Tr} (Y_L \rho) Y + 
    \text{Tr} (Z_L \rho) Z}{2}.
    \label{to_logical_subspace}
\end{equation}

The time dependence of the recovered state fidelity is presented in Fig.~\ref{dif_init_states}a. The data points are averages of upper bound fidelities in independent runs of the density matrix simulation with different ancilla measurement results. For each of the six initial states $|0\rangle, |1\rangle, |+\rangle$, $|-\rangle, |i+\rangle$ and $|i-\rangle$ 10000 simulations have been run.
The simulated points are approximated with the function \cite{OBrien2017}
\begin{equation}
    F(t) = \frac{1}{2} + \frac{1}{2} (1 - 2\epsilon)^{t-t_0},
    \label{fid_approx}
\end{equation}
where $\epsilon$ is the error rate per $1 \, \mu$s, $t$ is time in $\mu$s. The parameters $\epsilon$ and $t_0$ are obtained from a the least squares fit and their values are presented in Table~\ref{ub10_17_table}. As expected for the 5-qubit code, the state preservation fidelities of the six select states do not differ from each other significantly.

\begin{table}[b!]
    \centering
    \begin{tabular}{|l|c|c|}
        \hline
        \hline
        & \multicolumn{2}{c|}{Error rate $\epsilon$ in \% per $1 \, \mu$s} \\
        \hline
        Initial state & 5-qubit  & Surface-17 \\
        \hline
        $|0\rangle$ & $0.644 \pm 0.006$ & $0.604 \pm 0.005$\\
        $|1\rangle$ & $0.629 \pm 0.006$ & $0.621 \pm 0.005$\\
        $|+\rangle$ & $0.675 \pm 0.002$ & $0.466 \pm 0.007$\\ 
        $|-\rangle$ & $0.665 \pm 0.004$ & $0.442 \pm 0.004$\\ 
        $|i+\rangle$ & $0.638 \pm 0.003$ & $1.143 \pm 0.020$\\
        $|i-\rangle$ & $0.623 \pm 0.005$ & $1.123 \pm 0.015$\\
        Mean & $0.646 \pm 0.005$ & $0.703 \pm 0.009$\\
        \hline
        \hline
    \end{tabular}
    \caption{Upper bound error rates $\epsilon$ as obtained from least-squares fitting of fidelity upper bounds shown in Fig.~\ref{dif_init_states} with Eq.~\ref{fid_approx} for the six different initial states in the five-qubit QEC code and Surface-17. Errors estimates for $\epsilon$ are computed from the variance matrix of the least squares method. ``Mean'' error rate corresponds to the average over the six initial states.}
    \label{ub10_17_table}
\end{table}

The first point of each time dependence is excluded from the fit, as it clearly does not lie on the same curve as the others. The reason for the abrupt fidelity drop after the first cycle is that our proposed logical qubit initialization procedure is flawed. While it does prepare an eigenstate of the logical qubit's operators with only single-qubit gates, preparing a state that is also an eigenstate of the stabilizers requires a complex circuit, which is challenging to implement with high fidelity in the proposed device topology. As a result, during the first cycle the ancilla measurements randomly project the system into one of the stabilizer eigenstates, and actual error correction starts only from the second cycle.

If we numerically prepare an eigenstate of the error correction code instead of the described initialization sequence, the curve starts from the first cycle, as shown in the inset plot in Fig.~\ref{dif_init_states}a for the $|0\rangle_L$ and $|1\rangle_L$ states. The fitted error rates $\epsilon$ are $0.598 \pm 0.006$ and $0.588 \pm 0.009$\% per $1 \, \mu$s.

To evaluate how well the five-qubit code fares in compare to other stabilizer codes, we simulate Surface-17 under the same noise conditions (Appendix~\ref{appendix:Surface-17}). The time dependence of the recovered state fidelity for Surface-17 averaged over 5000 independent runs per each of the six initial states is presented in Fig.~\ref{dif_init_states}b and in Table~\ref{ub10_17_table}. In contrast to the five-qubit code it appears that different states are preserved with considerably different fidelity, which leads to worse preservation of arbitrary states on average.

\begin{figure}[t]
    \center{\includegraphics[width=\linewidth]{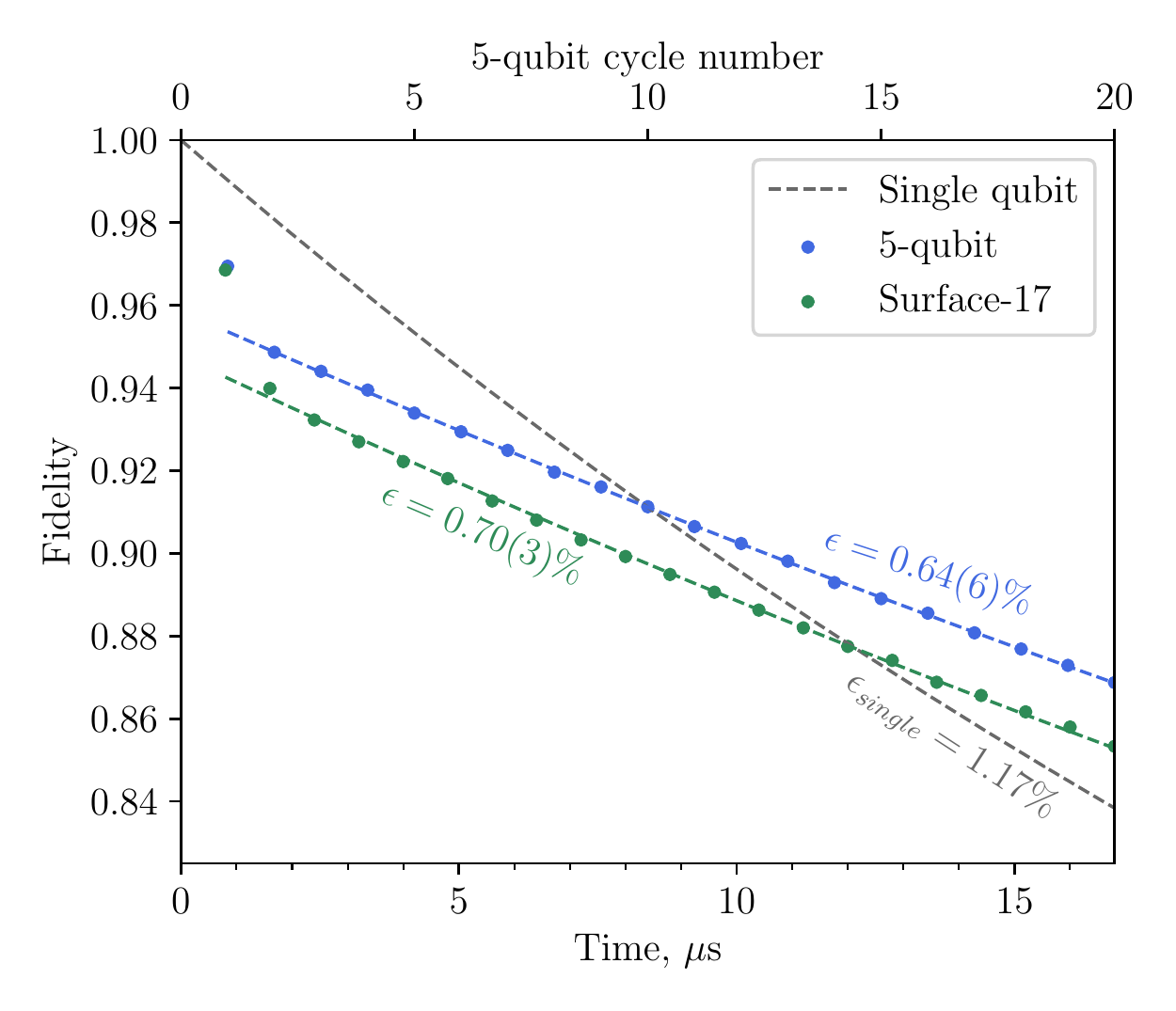}}
    \caption{Upper bound fidelities of the five-qubit (blue) and Surface-17 (green) error correction codes, averaged over the six initial states $|0\rangle, |1\rangle, |+\rangle$, $|-\rangle, |i+\rangle$ and $|i-\rangle$, calculated via the density-matrix simulation. The grey curve corresponds to the average state preservation fidelity in a single physical qubit.}
    \label{compare_ub}
\end{figure}

Also we compare the both correction codes to the single qubit relaxation with times $T_1$ and $T_2$. In this case the fidelity averaged over the six initial states is given by the formula
\begin{equation}
    F_{\text{single}}(t) = \frac{1}{6} \left( 1 + e^{-t/T_1} \right) + \frac{1}{3} \left( 1 + e^{-t/T_2} \right).
    \label{single_fid}
\end{equation}

In Fig.~\ref{compare_ub} we plot the mean state preservation fidelity of the considered QEC codes over the six cardinal states.

\begin{figure}[t]
    \center{\includegraphics[width=\linewidth]{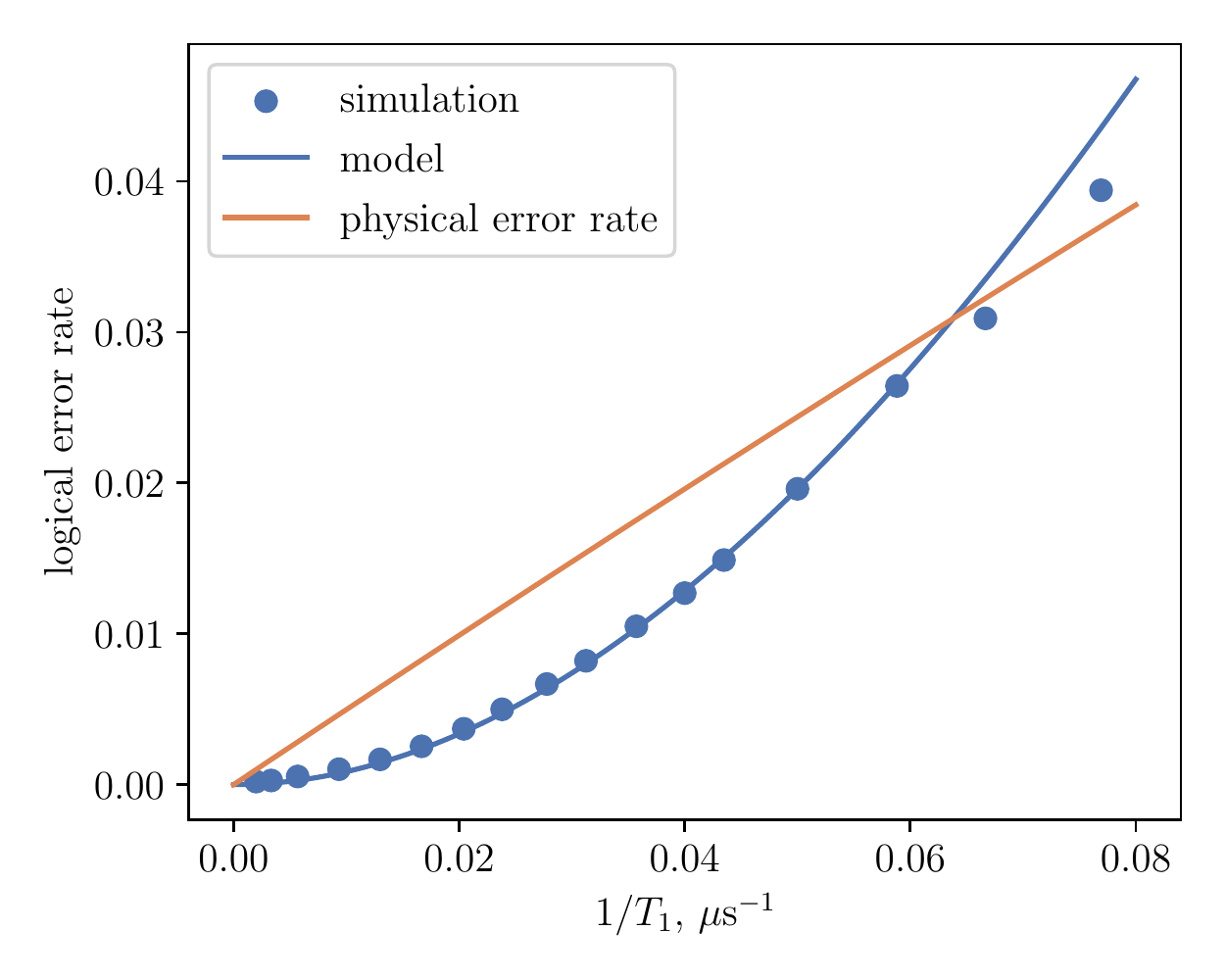}}
    \caption{ Upper bound logical error rate $\epsilon$ of the five-qubit code in presence of only relaxation and dephasing ($T_1=T_2)$, averaged over the three initial states $|0\rangle, |+\rangle$ and $|i+\rangle$. The blue points are data calculated via the density-matrix simulation; blue curve is their fit with an Eq.~\ref{simpleformula}; orange line corresponds to a physical error rate of a single qubit.}
    \label{fig:logicalerrorrate}
\end{figure}

A noteworthy question is the dependency of the logical error rate on the quality of the qubits. To study this issue, we simplify the error model and take into account only the decoherence of qubits described by the relaxation times $T_1$ and $T_2$. It is essential to note that qubits relax not only between correction cycles, when the ancillas are being measured but also during the whole correction procedure. We simulate the state preservation experiment for several times $T_1$ (the transverse relaxation time $T_2$ is chosen equal to $T_1$) and obtain the corresponding upper-bound logical error rate $\epsilon$) shown in Fig.~\ref{fig:logicalerrorrate}.

To approximate the data we suggest following simple model. Let $N$ qubits participate in the experiment and $p$ is an independent probability of a single-qubit error per unit of time. Then the probability that more than two single-qubit errors happen during a correction cycle can be estimated by the following expression:
\begin{equation}
    P = 1-\left[(1-p)^N + Np(1-p)^{N-1}\right],
    \label{simpleformula}
\end{equation}
that for small $p$ it can be expressed in the series:
\begin{equation}
    P = \frac{(N-1)N}{2}p^2+\mathcal{O}(p^3).
\end{equation}

Since the five-qubit code protects the logical state against one single-qubit error, then it is reasonable to approximate the dependence of the logical error rate on time $T_1$ by formula~\ref{simpleformula}, where $p$ is intuitively proportional to $p=\beta\exp(-t/ T_1)$ with some proportionality coefficient $\beta$. Optimizing this parameter $\beta$ ($\beta_\mathrm{fit}=0.44$) it turns out that the theoretical curve fits well the simulated values at $T_1 \geq 20 \mu$s. For smaller $T_1$ some divergence is observed, which is explained by the fact that the model described above can be valid only for small $p$.

In addition, comparing the formulas~\ref{fid_approx}~and~\ref{single_fid}, one can see that in case $T_1 = T_2$ logical error rate per $1 \mu$s for a single qubit is $\epsilon = 0.5\exp(-1/T_1)$  [$T_1$ in $\mu$s] and, therefore, consider it as a physical error rate.

\begin{figure*}[t]
    \center{\includegraphics[width=\linewidth]{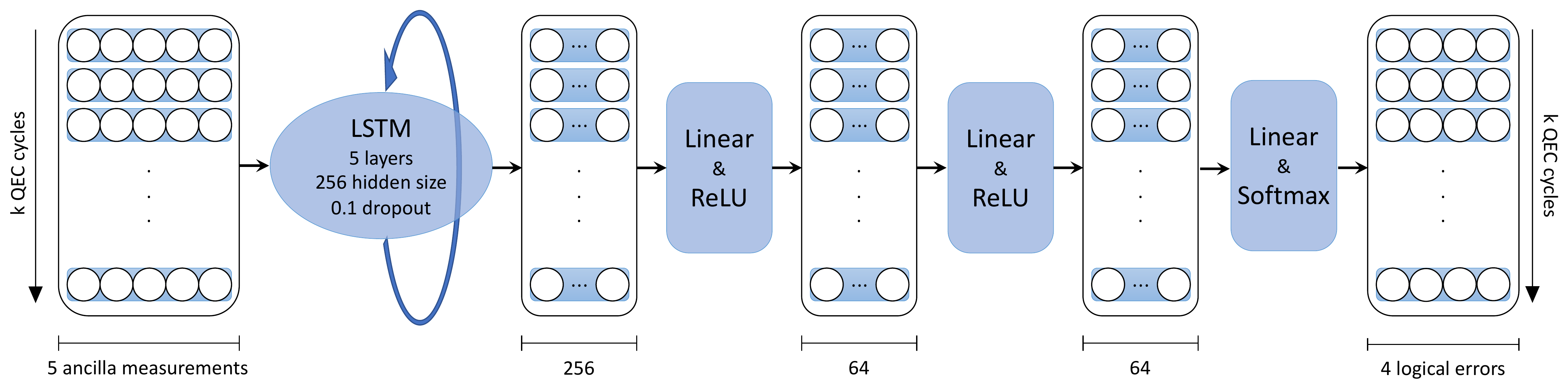}}
    \caption{Architecture of the neural network decoder. The network takes measurements of the five ancillary qubits during $k$ cycles as an input, processes them with 5 LSTM layers with hidden size 256, then passes them through a perceptron with 2 hidden layers of size 64, and output size of 4 neurons, and ReLU activation functions. After a final softmax activation function, the outputs are interpreted as probability predictions for each Pauli operator error $p(X), p(Y), p(Z)$ or $p(I)$.}
    \label{NNarchitecture}
\end{figure*}

\subsection{Neural network decoder}

The upper bound is the theoretical limit of any decoder performance for a given error model. Realizing this decoder amounts to simulating the system's evolution with this error model, which can be challenging for large systems and systems with complex error statistics.

Artificial neural network decoders are an alternative approach altogether. The problem of any decoder is to find and analyze patterns in the sequences of ancillary qubit measurements. Instead of programming an algorithm that would detect patterns that physical errors produce in the ancilla measurements, one can train a neural network to detect these patterns from a training dataset.
Due to the nature of the errors, data about the ancilla measurements from all previous steps may contain information about which logical error has which probability on each step. This time series-like type of feature data can be effectively accommodated by recursive neural network architectures, such as the Long Short-Term Memory (LSTM) \cite{lstmOrig}. The performance of LSTM-based neural network decoders has been studied for color codes \cite{baireuther2019neural} and surface codes \cite{baireuther2018machine}; the state preservation fidelity obtained with these decoders is on par or better than with more traditional decoders.

These inputs are fed into a five-layer LSTM network with a hidden layer size of 256 neurons Fig.~\ref{NNarchitecture}. The LSTM's outputs are passed on to a perceptron with two hidden layers with 64 neurons, an output layer of 4 neurons, and a ReLU activation function. After a final softmax activation layer, the neural network outputs are interpreted as probability predictions for each Pauli operator error $p(X), p(Y), p(Z)$ or $p(I)$ (no error). Our neural network implementation is based on the PyTorch framework.

A training dataset contains the simulated ancilla measurement results and readout results of the final data qubit states. The neural network is trained simultaneously with three training datasets with different initialization and projection operators $P$ corresponding to three different cardinal points of the Bloch sphere $\{|+\rangle, |i+\rangle, |0\rangle\}$, with $4\times 10^6$ readout results in each dataset.

From formula~\eqref{eq:logical_operators} it follows that the readout result of the logical qubit's Pauli operators $\Pi_L \in \{X_L, Y_L, Z_L\}$ can be expressed as 
\begin{equation}
    \Pi_L = \prod\limits_{i=0}^4\Pi_i.
\end{equation}
If $\Pi_L = +1$, then effectively the qubit has not flipped, and the logical error operator is either $I$ or $\Pi$. Otherwise, it is one of the other two Pauli operators. We define a loss function based on the cross-entropy of the experimentally measured logical error operator and the neural network outputs over the set of logical error operators as

\begin{multline}
    L = \frac{1}{2}
    \begin{cases}
        \log p(I) + \log p(\Pi),& \text{ if } \Pi_L = +1,  \\ \sum\limits_{\mathcal{P} \in \{X, Y, Z\} / \{\Pi\}}\log p(\mathcal{P}),& \text{ otherwise}.
    \end{cases}
\end{multline}

\begin{figure}
    \center{\includegraphics[width=\linewidth]{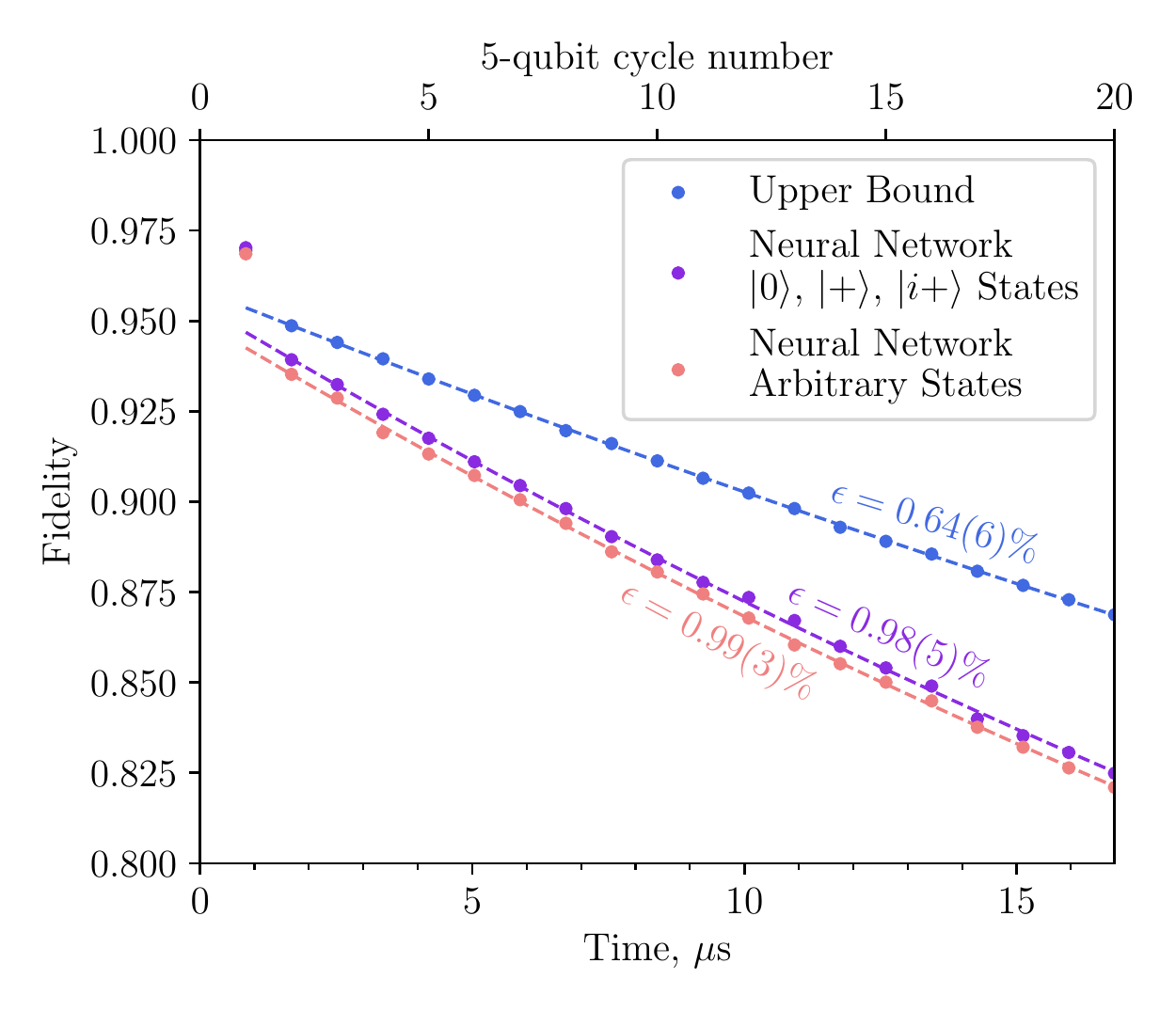}}
    \caption{Comparison of recovered logical qubit state fidelity in the five-qubit QEC code for a perfect decoder (upper bound) and the neural network decoder applied to $|0\rangle, |+\rangle, |i+\rangle$ initial states and to arbitrary uniformly distributed pure states. }
    \label{NN_performance}
\end{figure}

The decoder performance is tested on 150000 runs from the same initial states, but as opposed to the training procedure we do not measure the final logical state but just calculate it from the density matrix of data qubits via the formula~\ref{to_logical_subspace} for better comparison with the upper bound decoder. The fidelity of the neural network decoder-recovered state is shown in Fig.~\ref{NN_performance} (violet curve) and the fitted error rate is given in Table~\ref{NN_table}.

\begin{figure}
    \center{\includegraphics[width=0.7\linewidth]{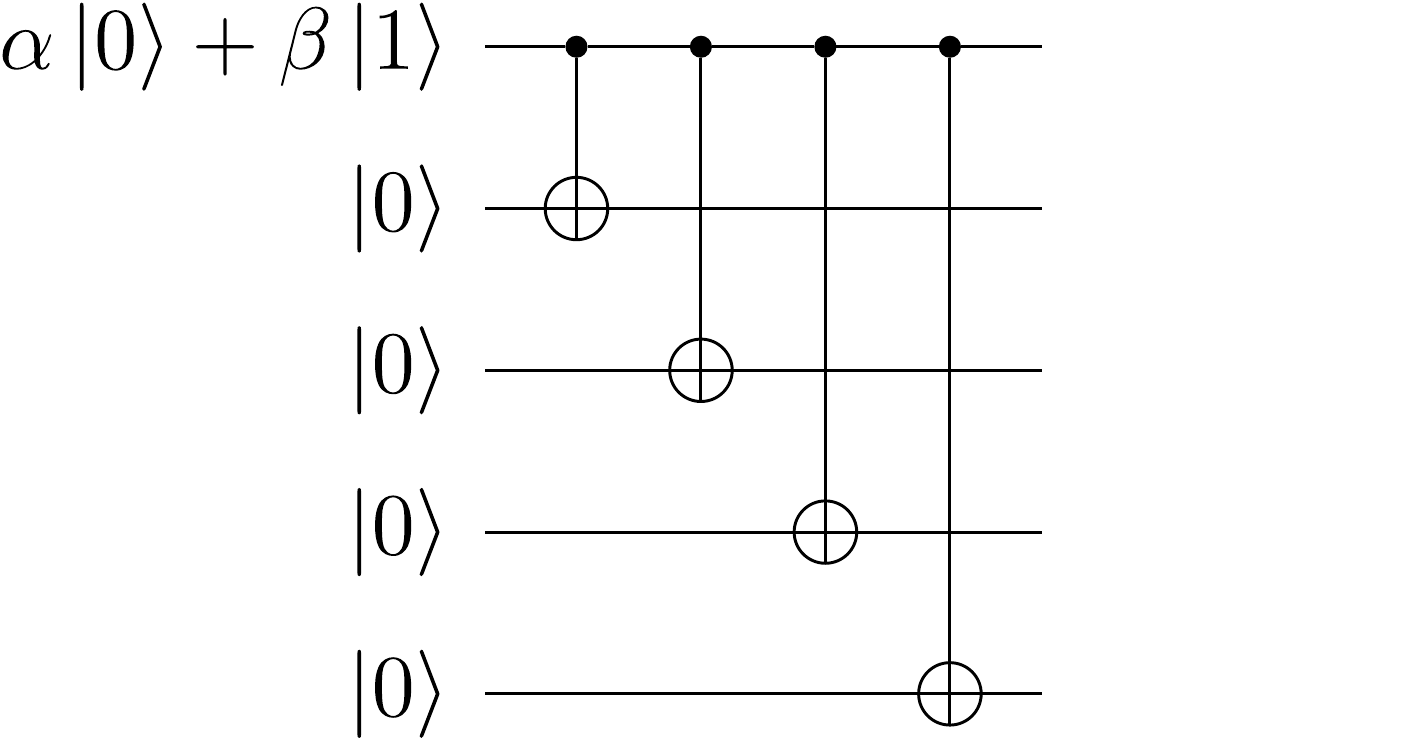}}
    \caption{The quantum circuit for arbitrary initial state preparation consisting of four CNOT gates that encodes $\alpha|0\rangle+\beta|1\rangle$ state into $\alpha|00000\rangle+\beta|11111\rangle$.}
    \label{encoding}
\end{figure}

However, the goal is to correct not only the six cardinal points on the Bloch sphere but an arbitrary state. To check how well the decoder fares in this case we generate a random pure single-qubit state $\alpha |0\rangle + \beta |1\rangle$, uniformly distributed on the Bloch sphere, and encode it into the state $\alpha |00000\rangle + \beta |11111\rangle$ by the circuit shown in Fig.~\ref{encoding}. We assume that the encoding circuit is error-free. After initialization, we simulate the state preservation experiment and apply the neural network decoder to the final density matrix. We repeat the whole procedure for 17000 different initial states and plot the average fidelity in Fig.~\ref{NN_performance} (pink curve). Taking into account the error, we observe approximately the nearly same $\epsilon$ as for the cardinal points (see Table~\ref{NN_table}).

\begin{table}
    \centering
    \begin{tabular}{|l|c|}
        \hline
        \hline
        & $\epsilon$, \% per $1 \, \mu$s \\
        \hline
        Initial state & Neural Network Decoder \\
        \hline
        $|0\rangle$ & $0.88 \pm 0.01$ \\
        $|+\rangle$ & $1.03 \pm 0.01$ \\
        $|i+\rangle$ & $1.05 \pm 0.01$ \\ 
        Mean over & \\ 
        $|0\rangle, |+\rangle, |i+\rangle$ states & $0.99 \pm 0.01$ \\ 
        Arbitrary states & $0.993 \pm 0.003$ \\
        \hline
        \hline
    \end{tabular}
    \caption{Error rates $\epsilon$ obtained with least-squares fitting of the fidelity time dependence for the neural network decoder shown in Fig.~\ref{NN_performance} for different initial states. The error is computed via the k-fold cross-validation method with $k=5$.}
    \label{NN_table}
\end{table}

\begin{table}[h!]
    \centering
    \begin{tabular}{|l|l|l|}
        \hline
        \hline
        Parameter & Symbol  & Value \\
        \hline
        Qubit relaxation time & $T_1$ & $30$ $\mu \text{s}$\\
        Qubit dephasing time & $T_2$ & $40$ $\mu \text{s}$\\
        Hadamard gate duration & $\tau_1$ & $20$ ns \\ 
        iSWAP gate duration & $\tau_{2}$ & $40$ ns \\ 
        Measurement time & $t_m$ & $300$ ns \\
        Reset time & $t_r$ & $300$ ns\\
        5-qubit cycle time & $t_c$ & $840$ ns \\
        \hline
        \hline
    \end{tabular}
    \caption{Time parameters used in the density-matrix simulation \cite{measerrorcite, timeerrorcite}.}
    \label{time_params}
\end{table}

\section{Conclusion}

In conclusion, we propose a circuit that implements the five-qubit quantum error correction code on a realistic topology of superconducting qubits. The key advantage of the scheme is that it requires only 10 qubits connected in a ring with two-qubit iSWAP gates available between any two neighboring qubits. The small footprint of our circuit makes it one of the least hardware-demanding proposals for stabilizer-based error correction. Using a density-matrix simulation of the state preservation experiment under a realistic noise model, we obtain estimates for logical error rates. We propose a decoder based on an artificial neural network, that can be trained on experimentally available data, which shows satisfactory performance close to the upper bound for preservation of arbitrary logical states.

\section*{Acknowledgements}

This work was performed with the financial support of the Russian Science Foundation, Project No. 21-72-30026.

\section*{Code availability}

The computer code developed for simulations is available from the corresponding author (\href{mailto:simakov.ia@phystech.edu}{simakov.ia@phystech.edu}) upon request.

\section*{Author contributions}

I.A.S. performed simulations and calculations. I.S.B. proposed the stabilizer circuit and the neural network decoder architecture. A.V.U. supervised the project. All authors contributed to the manuscript.

\renewcommand{\bibname}{Reference}
\bibliographystyle{unsrtnat}
\bibliography{main}
    

\newpage
\newpage
\appendix

\section{Density-matrix simulation details}

In order to estimate the limit of the QEC code performance we need to simulate the dynamics of the investigated open quantum system, described with the density matrix $\rho$.
The time evolution $\rho \rightarrow \mathcal{E}(\rho)$ can be modeled with the operator-sum representation \cite{NielsenChuang}:
\begin{equation}
    \mathcal{E}(\rho)=\sum_{i} E_{i} \rho E_{i}^{\dagger},
    \label{operator-sum_rep}
\end{equation}
where $E_i$ are Kraus operators:
\begin{equation}
    \sum_{i=0}^n E_i^\dagger E_i = I.
\end{equation}

For the convenience of calculations we use the Pauli transfer matrix representation of the superoperator $\mathcal{E}$:
\begin{equation}
    (R_\mathcal{E})_{ij} = \frac{1}{2} \text{Tr} \left( \sigma_i \mathcal{E} \left(\sigma_j \right)\right),
\end{equation}
where $\sigma_0$ is the identity matrix $I$ and $\sigma_1, \sigma_2, \sigma_3$ are Pauli matrices $X, Y, Z$.
In a such form the time evolution, including quantum gates and noise channels, can be expressed as simple multiplication of the transfer matrix $(R_\mathcal{E})_{ij}$ and vector representation of the density matrix in the Pauli basis.

The Hilbert space of the quantum system implementing the QEC code (Fig.~\ref{circuit-10}) consists of 10 qubits.
The density matrix of the system is represented as a vector of dimension $4^{10}$ and transfer matrices, acting on this vector, have the size $4^{10} \times 4^{10}$. In order to accelerate calculation we use sparse matrices, batches and GPUs.

\section{Error models}
\label{appendix:errors}

This appendix contains the mathematical description of the noise channels that we take into account in the density-matrix simulation. The basic motivation of all the error models introduced in the paper is to get the simulation closer to real experiments \cite{measerrorcite, timeerrorcite, fluxnoisecite1, fluxnoisecite2}. The values of the error model parameters are given in Table \ref{error_params_full}, excluding measurement errors which are listed separately.

\begin{table}[h!]
    \begin{center}
        \begin{tabular}{|l|l|l|}
            \hline
            \hline
            Parameter & Symbol  & Value \\
            \hline
            Qubit relaxation time & $T_1$ & $30$ $\mu \text{s}$\\
            Qubit dephasing time & $T_2$ & $40$ $\mu \text{s}$\\
            single-qubit gate duration & $\tau_1$ & $20$ ns \\ 
            two-qubit gate duration & $\tau_{2}$ & $40$ ns \\ 
            Measurement time & $t_m$ & $300$ ns \\
            Reset time & $t_r$ & $300$ ns \\
            In-axis rotation error & $p_\text{axis}$ & $10^{-4}$ \\
            In-plane rotation error & $p_\text{plane}$ & $5 \cdot 10^{-4}$ \\
            Root-mean-square error & $\phi_\text{rms}$ & $0.01$ \\
            \hline
            \hline
        \end{tabular}
    \end{center}
    \caption{Parameters of error models \cite{measerrorcite, timeerrorcite, fluxnoisecite1, fluxnoisecite2}.}
    \label{error_params_full}
\end{table}

\subsection{Idling}

In the simulation we use the standard model of amplitude and phase damping channels.
The operation elements of these channels are
\begin{equation}
    E^\text{AD}_0=\left(\begin{array}{cc}
    1 & 0 \\
    0 & \sqrt{1-\gamma_1}
    \end{array}\right), \quad E^\text{AD}_1=\left(\begin{array}{cc}
    0 & \sqrt{\gamma_1} \\
    0 & 0
    \end{array}\right)
\end{equation}
and
\begin{equation}
    E^\text{PD}_0=\left(\begin{array}{cc}
    1 & 0 \\
    0 & \sqrt{1-\gamma_\phi}
    \end{array}\right), \quad E^\text{PD}_1=\left(\begin{array}{cc}
    0 & 0 \\
    0 & \sqrt{\gamma_\phi}
    \end{array}\right),
\end{equation}
where $\gamma_1 = 1 - e^{-t/T_1}$, $\gamma_\phi = 1 - e^{-2t/T_\phi}$, $\frac{1}{T_2} = \frac{1}{2T_1} + \frac{1}{T_\phi}$. Corresponding Pauli transfer matrices are
\begin{equation}
    R_{\mathcal{E}_\text{AD}}=\left(\begin{array}{cccc}
    1 & 0 & 0 & 0 \\
    0 & \sqrt{1-\gamma_{1}} & 0 & 0 \\
    0 & 0 & \sqrt{1-\gamma_{1}} & 0 \\
    \gamma_{1} & 0 & 0 & 1-\gamma_{1}
    \end{array}\right),
\end{equation}
\begin{equation}
    R_{\mathcal{E}_{\text{PD}}}=\left(\begin{array}{cccc}
    1 & 0 & 0 & 0 \\
    0 & \sqrt{1-\gamma_{\phi}} & 0 & 0 \\
    0 & 0 & \sqrt{1-\gamma_{\phi}} & 0 \\
    0 & 0 & 0 & 1
    \end{array}\right).
\end{equation}

Thus, the evolution of an idling qubit for the duration $t$ is described by transfer matrix
\begin{equation}
    R_{\text{idle}}(t) = R_{\mathcal{E}_{\text{AD}}} R_{\mathcal{E}_{\text{PD}}}.
\end{equation}

\subsection{Single-qubit gates}

There are three single-qubit gates in the circuit Fig.~\ref{circuit-10}: Hadamard ($H$), Phase ($S$) and Pauli-Z ($Z$). The last two gates are considered to be virtual, instantaneous and thereby ideal.
The Hadamard gate can be decomposed into rotations around the $z$-axis and the $y$-axis: $H = Z R_y(-\pi/2)$. 
Thereby, it is necessary to build an error model for the $y$-axis rotation.
This model should include amplitude and phase damping, depolarizing and statistical error of the angle of rotation.

Processes of amplitude and phase decoherence are taken into account with the sandwich model, where the gate transfer matrix is expressed as the product of the two idling process matrices interleaved with the ideal gate process matrix.

In order to have better conformity with tomography experiments, we additionally introduce a depolarization channel, that geometrically corresponds to squeezing along the $x$ and $z$ axes with the factor $1-p_\text{plane}$ and the $y$ axis with the factor $1-p_\text{axis}$.
The depolarization Pauli transfer matrix is
\begin{equation}
    R_{\mathrm{dep}}=\left(\begin{array}{cccc}
    1 & 0 & 0 & 0 \\
    0 & 1-p_{\text {plane }} & 0 & 0 \\
    0 & 0 & 1-p_{\text {axis }} & 0 \\
    0 & 0 & 0 & 1-p_{\text {plane }}
    \end{array}\right).
\end{equation}

Finally, we include the statistical error for the angle of rotation $\phi$.
We consider $\phi$ to be normally distributed with mean $\varphi_0$ and variance $\sigma^2$. The expectation value of the exponential function is
\begin{equation}
   \langle \mathrm{e}^{i \varphi} \rangle = \int\limits_{-\infty}^{\infty} \frac{1}{\sqrt{2\pi}\sigma}  \mathrm{e}^{-\frac{(\varphi - \varphi_0)^2}{2\sigma^2}} \mathrm{e}^{i \varphi} \mathrm{d} \varphi= \mathrm{e}^{i \varphi} \mathrm{e}^{-\frac{\sigma^2}{2}}.
\end{equation}
Using this result we add stochastic error to the matrix elements of the rotation gates. These stochastic errors can be attributed to flux noise and long-term drift of microwave signal source phase and amplitude. Phenomenological estimates yield an RMS value in the order of $\sigma = \varphi_{\text{rms}} \approx 0.01$ rad.

According to the model above, the Hadamard gate can be expressed as
\begin{equation}
    R_H = R_{\text{idle}}\left(\frac{\tau_1}{2}\right) R_{\mathrm{dep}} R_{R_z(\pi)} R'_{R_y\left(-\frac{\pi}{2}\right)} R_{\text{idle}}\left(\frac{\tau_1}{2}\right),
\end{equation}
where $\tau_1$ is the single-qubit gate duration and
$R'_{R_y\left(-\frac{\pi}{2}\right)}$ is the PTM of the rotation gate with the stochastic error. 

\subsection{Two-qubit gates}

The only two-qubit gate used in the present five-qubit code is an iSWAP gate.
This gate is generated by XY-type interaction, which is natural for many platforms for quantum computing, including superconducting qubits  \cite{CZiSWAPnoise, schuch2003natural}, and the unitary matrix of such interaction is
\begin{equation}
    U(\theta, \eta, \zeta)=\left(\begin{array}{cccc}
    1 & 0 & 0 & 0 \\
    0 & \cos \theta / 2 & i e^{i \eta} \sin \frac{\theta}{2} & 0 \\
    0 & i e^{-i \eta} \sin \theta / 2 & \cos \theta / 2 & 0 \\
    0 & 0 & 0 & e^{i\zeta}
    \end{array}\right).
\end{equation}
For iSWAP the parameters are $U_{\mathrm{iSWAP}} = U(\pi, 0, 0)$. Here we assume the same stochastic error independent and equal for both parameters $\theta$ and $\eta$ as it is described in the section above for the single-qubit rotation gate. The parameter $\zeta$ takes into account the conditional phase arising in the dispersive limit due to the $|11\rangle \leftrightarrow |20\rangle$ interaction in transmon qubits. The error for $\zeta$ is correlated with the single-qubit stochastic RMS deviation due to flux noise, such that $\sigma_\zeta = \varphi_{\text{rms}}/2 = 0.005$~rad.

The Surface-17 code uses the controlled-Z (CZ) gate with the following unitary matrix: $U_{\text{CZ}} = U(0, 0, \pi)$. For CZ we take into account stochastic error only for parameter $\zeta$.

Decoherence is modeled by a sandwich structure, as it was described above for single-qubit rotation gates, and the final PTM of the two-qubit gate $G$ is
\begin{equation}
    R_{G} = R_{\text{idle}}\left(\frac{\tau_2}{2}\right) R'_G R_{\text{idle}}\left(\frac{\tau_2}{2}\right).
\end{equation}
Here $\tau_2$ is the duration of the gate, and $R'_G$ is the PTM of the two-qubit gate with stochastic errors.

\subsection{Measurement}

The measurement procedure is realized by the method, described below, in order to approximated the experimental results obtained in \cite{measerrorcite}. We use the model of back-to-back measurements. Firstly, we calculate from the density matrix the probabilities of a qubit to be projected in $|0\rangle$ and $|1\rangle$ states and then randomly with  respect to these probabilities choose the initial state $i \in \{|0\rangle, |1\rangle \}$. The second step is to determine the measurement result $m \in \{+1, -1\}$ and outcome state $o \in \{|0\rangle, |1\rangle \}$ of the qubit after the measurement procedure. We match with initial and outcome states according to the experimentally obtained probabilities given in the Table \ref{meas_prob}.

\begin{figure*}[t]
    \centering
    \includegraphics[width=\linewidth]{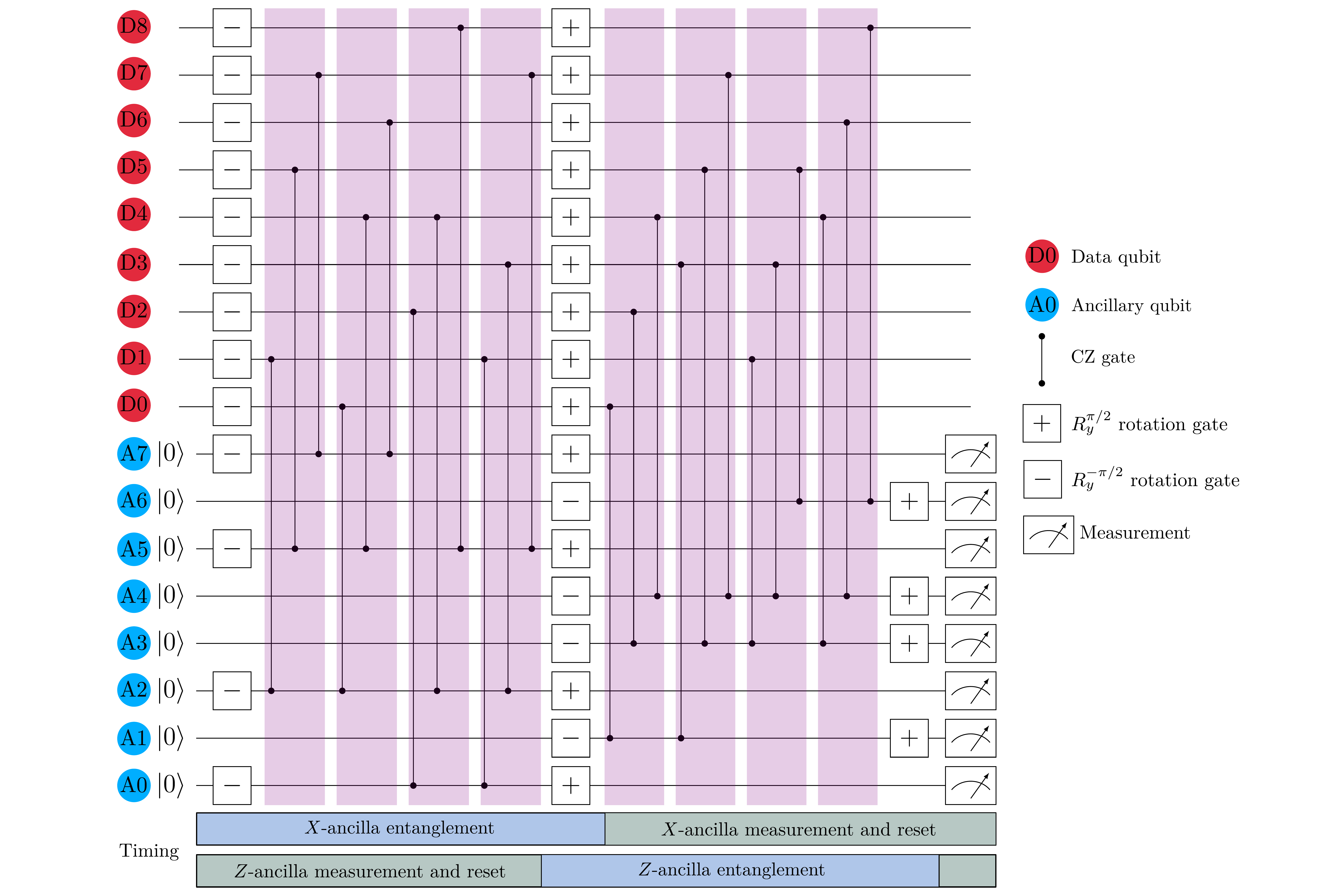}
    \caption{Quantum circuit of the QEC cycle of Surface-17 code, consisting of single-qubit rotations about the $y$-axis and eight series of simultaneous CZ gates.}
    \label{fig:surface17circuit}
\end{figure*}

\begin{table}[h!]
    \begin{center}
        \begin{tabular}{|l|l|l|l|}
            \hline \hline \text { Probability } & \text { Value } & \text { Probability } & \text { Value } \\
            \hline $\epsilon_{0}^{+1,0}$ & $0.9985$ & $\epsilon_{1}^{+1,0}$ & $0.0050$ \\
            $\epsilon_{0}^{+1,1}$ & $0.0000$ & $\epsilon_{1}^{+1,1}$ & $0.0015$ \\
            $\epsilon_{0}^{-1,0}$ & $0.0015$ & $\epsilon_{1}^{-1,0}$ & $0.0149$ \\
            $\epsilon_{0}^{-1,1}$ & $0.0000$ & $\epsilon_{1}^{-1,1}$ & $0.9786$ \\
            \hline \hline
        \end{tabular}
    \end{center}
    \caption{Probabilities of different errors in measurement procedure \cite{OBrien2017, measerrorcite}.}
    \label{meas_prob}
\end{table}

\section{Surface-17}
\label{appendix:Surface-17}

The simulated experimental procedure for the surface code simulation is similar to the five-qubit code simulation procedure. It can be divided into three stages: data qubit preparation, correction cycles and data qubit measurement. The first and the third stages are the same as in the five-qubit code with the only difference being the amount data qubits. The quantum circuit for the QEC cycle of Surface-17 is presented in Fig.~\ref{fig:surface17circuit} \cite{OBrien2017}. It consists of single-qubit rotations about $y$-axis and four series of simultaneous CZ gates (three gates per series). The syndromes for Surface-17 are
\begin{equation}
    \begin{aligned}
    S_{A0} &= X_1 X_2\\
    S_{A1} &= Z_0 Z_3\\
    S_{A2} &= X_0 X_1 X_3 X_4\\
    S_{A3} &= Z_1 Z_2 Z_4 Z_5\\
    S_{A4} &= Z_3 Z_4 Z_6 Z_7\\
    S_{A5} &= X_4 X_5 X_7 X_8\\
    S_{A6} &= Z_5 Z_8\\
    S_{A7} &= X_6 X_7,
    \end{aligned}
\end{equation}
where the indices of Pauli $X$ and $Z$ operations relate to the number of data qubit.

It is remarkable that $X$-ancilla measurement starts right after the $R_y^{\pi/2}$ gate on qubits \{A0, A2, A5, A7\}, and $Z$-ancilla measurement continues during the first four series of CZ gates as it is shown in the timeline in Fig~\ref{fig:surface17circuit}. Thereby, the full correction cycle time consists of 4 two-qubit gates, 2 single-qubit rotations, 1 measurement and reset procedure, totaling at 800~ns. 

The circuit is simulated under the same noise models as the five-qubit code.

\end{document}